\documentclass[fleqn,twoside]{article}
\usepackage{espcrc2}
\usepackage{graphicx}
\usepackage{subfigure}

\title{Rough Gauge Fields, Smearing and Domain Wall Fermions}
\author{Meifeng Lin
  \address{Department of Physics, Columbia University, New York, NY 10027,USA}
	  [RBC Collaboration]
	  \thanks{We thank RIKEN, Brookhaven National Laboratory and the U.S.\ Department
	    of Energy for providing the facilities essential for the completion of
	    this work.}}

\begin{document}
  
  \begin{abstract}
    At a fixed lattice spacing, as determined by say $m_\rho$, adding
    additional fermion flavors to a dynamical simulation produces rougher
    gauge field configurations at the lattice scale.  For domain wall
    fermions, these rough configurations lead to larger residual chiral
    symmetry breaking and larger values for the residual masses,
    $m_{\rm res}$.  We discuss ongoing attempts to reduce chiral symmetry
    breaking for $N_f = 3$ dynamical domain wall fermion simulations by
    different smoothing choices for the gauge fields.
  \end{abstract}
  
  \maketitle

  \section{INTRODUCTION}
  \label{sec:intro}
  As is widely discussed (see, for example, Ref.\cite{ref:localization}),
  topological lattice dislocations give rise to a non-vanishing density of near-zero modes of the hermitian 
  Wilson-Dirac operator (HWDO), which is considered to be an origin of large chiral symmetry breaking  for domain wall fermions (DWF) and a sign of 
  the theory being close to or inside the Aoki phase.   
  In quenched simulations, we are able to suppress the creation of these lattice dislocations 
  by using improved gauge actions \cite{ref:DBW2} (DBW2 for instance) and have achieved better chiral symmetry in numerical simulations. 
  However, if the lattice spacing, as determined by, say, the $\rho$ mass,
  is kept fixed as more flavors of fermions are added, asymptotic freedom
  predicts a rougher gauge field at the single-lattice spacing scale.
  As a consequence of the
  rougher fields, the residual mass  $m_{\rm res} $\cite{ref:quenchedDWF},
  becomes
  larger at a comparable lattice spacing than in the quenched theory.
  
  Recently there have been various attempts \cite{ref:improve} to improve the chiral properties of DWF in dynamical simulations by modifying the gauge action.
  Here we study a smearing technique on $N_f=3$ lattices,
  \textit{i.e.}, replacing each gauge link by a 
  weighted sum of the original link and products of gauge matrices along 3, 5
  and 7 link paths, to see if chiral symmetry breaking is reduced.
  
  \section{SMEARING ON GAUGE LINKS}
  \label{sec:smearing}
  A smeared link is constructed as:
  \vspace{-2pc}
  \begin{figure}[ht]
    \includegraphics[width=0.45\textwidth]{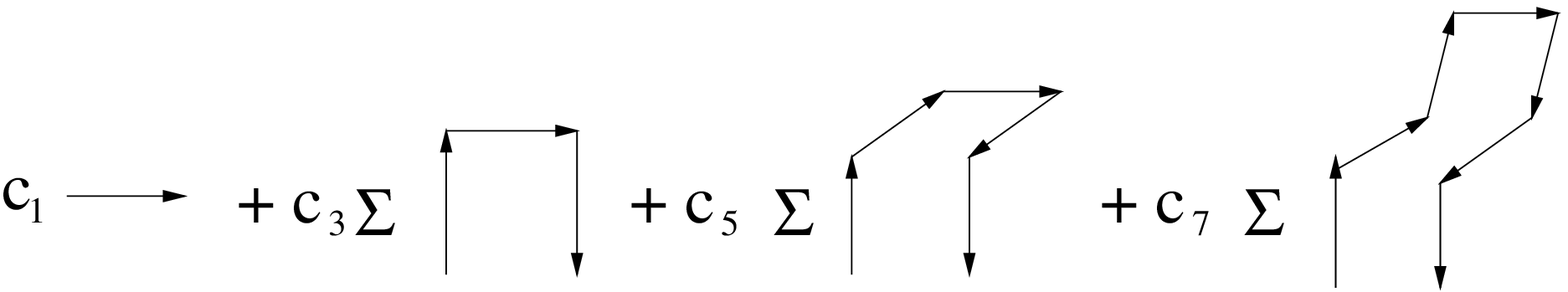}
    \label{fig:smear}
  \end{figure}
  \vspace{-2pc}

  \noindent As we can see, the new links are not necessarily elements of $SU(3)$.
  Although this may make the interpretation
  of the unrenormalized observables obscure, it does have the advantages of  cheap computer time cost and
  simple numerical implementation in full QCD simulations. 
  Yet the $c$'s should be appropriately
  chosen so that the conventional field normalization in weak coupling is
  preserved.
  
  The $c$'s can be simply normalized as: 
  \begin{equation}
    c_1+6c_3+24c_5+48c_7=1
    \label{eq:weakl}
  \end{equation}
  While this preserves the field normalization in weak coupling, at the lattice
  spacings of current simulations a better choice has
  been seen to be the so-called tadpole improvement condition, in which the smearing coefficients are chosen so that
  \begin{equation}
    c_1+6c_3 u_0^2+24c_5 u_0^4+48c_7 u_0^6=1
    \label{eq:tad-imp}
  \end{equation}   
  where $u_0$ is the quartic root of the average plaquette.  We explore these
  and other normalization possibilities 
  by measuring $m_{\rm res}$ with a fixed $M_5=1.8$ for different combinations of
  $c_1$ and $c_3$ without enforcing a pre-defined normalization. Coefficients which give smaller
  $m_{\rm res}$ are then further studied.   
  
  \section{SIMULATION RESULTS} 
  
  \subsection{Residual masses}
  \label{sec:mres}
  Smearings are performed on our $16^3\times32$, $N_f=3$ lattices, which
  were generated with the DWF and DBW2 gauge action. The parameters used
  were $\beta=0.72$, $M_5=1.8$, $L_s=8$ and $m_{\rm sea}=0.04$,
  yielding a lattice spacing (from $m_\rho$) of $a^{-1} \approx 1.6$ to 1.7
  GeV \cite{ref:3flavor}. Choices for the smearing coefficients we have studied include (i) one set normalized according to eq.(\ref{eq:weakl}):
  $c_1={1\over8}, c_3={1\over16}, c_5={1\over64}, c_7={1\over384}$,\textit{a la} MILC's Fat7; 
  (ii) one set following eq.(\ref{eq:tad-imp}): $c_1=0.25, c_5=0.051$;
  (iii)\&(iv) two sets with arbitrary normalizations for comparison: $c_1=0.4, c_3=0.12$ and $c_1=0.8, c_3=0.06$. 
  In Table~\ref{table:mres}, we see that $m_{\rm res}$ has generally decreased 
  for different choices of smearing coefficients except set (i).
  \begin{table}[t]
    \caption{ $m_{\rm res}$ with different smearings. All the measurements
      have valence $L_s=8$ and $M_5=1.8$ except the one with $c_7 \ne 0$, which has 
      $M_5=2.2$. Data for unsmeared lattices are from 84 configurations; others are all from 30 configurations.}
    \begin{tabular}{l@{\extracolsep{1mm}}lllcl}
      \hline
      $c_1$ & $c_3$ & $c_5$ & $c_7$ & $m_{\rm val}$ & $m_{\rm res}$ \\
      \hline
      \hline
      1.0 &   &   &  &  0.02  & 0.01135(7) \\
      &   &   &  &  0.03  & 0.01115(6) \\
      &   &   &  &  0.04  & 0.01097(6) \\
      &   &   &  &  0.05  & 0.01083(5) \\
      \hline
      ${1\over8}$ & ${1\over16}$ & ${1\over64}$ & ${1\over384}$ & 0.02 & 0.04811(28) \\
      &       &        &       & 0.04 & 0.04633(27) \\  
      
      \hline
      0.25  &      &0.051&  & 0.02  & 0.00651(13) \\
      &      &     &  & 0.04  & 0.00609(11) \\
      \hline
      0.4 & 0.12 &   &   &  0.02  & 0.00904(14)  \\
      &      &   &   &  0.04  & 0.00855(12)  \\  
      \hline
      0.8 & 0.06 &   &   &  0.02  & 0.00580(11)  \\
      &      &   &   &  0.04  & 0.00551(9)   \\
      \hline
      \hline
    \end{tabular}
    \vspace{-1.5pc}
    \label{table:mres}
  \end{table}
  \begin{figure}[t]
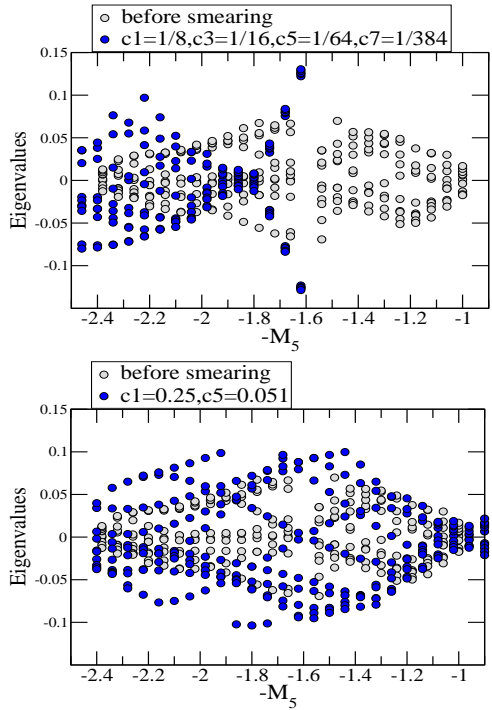

    \centering
    \subfigure{
      \includegraphics[height=11pc,width=0.40\textwidth]{figs/gf-cn1.eps}
    }\\
    \vspace{-1.5pc}
    \subfigure{
      \includegraphics[height=11pc,width=0.40\textwidth]{figs/gf-tad5.eps}
    }\\
    \vspace{-3pc}
    \caption{Eigenspectra of the hermitian Wilson-Dirac operator for different smearing coefficients on the same lattice. 
      \vspace{-1pc}
    }
    \label{fig:sflow}
  \end{figure}

  For DWF, the value of $m_{\rm res}$ is affected by the value for $M_5$.
  However, smearing can change the normalization of terms in the action which
  involve the gauge field, relative to the normalization of the mass term $M_5$.
  We can see this by considering
  eigenspectra of the HWDO (spectral flows) on the smeared gauge configurations.
  Shown in Figure~\ref{fig:sflow} are representative spectral flows from smeared and unsmeared lattices\footnote[1]{The author thanks \emph{S.D.Cohen} for providing the data necessary for this figure.}. The pinch point ($\kappa_c$ for Wilson fermions) on the top graph (Fat7 smearing) has shifted from $-M_5\sim -0.9$ to $\sim -1.8$. The larger $m_{\rm res}$ for this smearing choice is a result of non-optimal $M_5$ and possibly rougher gauge configurations.     
  On the bottom graph (tadpole smearing), the density of small eigenvalues is visibly decreased and the gap is widely opened near $-M_5=-1.8$,  
  leading to the smaller residual masses with coefficient choice (ii).  
  However, it is also worth noting
  that the major crossing at $-M_5\sim -1.6$ still persists after smearing. 
  The failure to eliminate these crossings indicates that the topological lattice dislocations are not changed.   While $m_{\rm res}$ has decreased for
  this $L_s$, the persistence of the topological dislocations
  makes studying the large-$L_s$ behavior of $m_{\rm res}$ important.  We
  present results on this in Sec.~\ref{sec:decay}.
  
  \subsection{Lattice scales and mass renormalization}
  \label{sec:scale-and-renorm}
  To solidify our conclusions about the effects of smearing on $m_{\rm res}$,
  we have studied the lattice spacing on smeared lattices using $m_\rho$,
  since it is known that $m_{\rm res}$ is strongly dependent on lattice
  spacing.  We determine the lattice scales by taking $m_\rho a$ in the valence
  quark chiral limit.  
  As shown in Table~\ref{table:mrho}, $m_\rho a$ is essentially unaffected by smearings.   
  
  Another important effect when comparing $m_{\rm res}$ with different
  smearing choices is the change in mass renormalization introduced by smearing. 
  Noting the fact that $m_\pi^2\propto m_q$, where $m_q=m_{\rm val}+m_{\rm res}$ for DWF, we can extract the smeared lattice mass
  renormalization factors relative to those on the unsmeared lattices by
  \begin{equation}
    Z_m={{{(m_\pi^s)}^2/m^s_q}\over{{(m_\pi^u)}^2/m^u_q}}
  \end{equation}  
  where the quantities on unsmeared lattices are superscripted by $u$ and
  those on smeared lattices by $s$. The renormalized residual masses are thus
  $m^r_{\rm res}=Z_m m_{\rm res}$. The $Z_m$'s are also shown in Table~\ref{table:mrho}.
  \begin{table}[t]
    \caption{ $m_\rho a$ in valence quark chiral limit and mass renormalization.
      $m_{\rm val}=0.02,0.03,0.04,0.05$ are used in the extrapolation for unsmeared lattices; 
      $m_{\rm val}=0.02,0.04$ are used otherwise.}
    \begin{tabular}{l@{\extracolsep{1mm}}lllll}
      \hline
      $c_1$ & $c_3$ & $c_5$ & $c_7$ & $m_{\rho}a$ & $Z_m$  \\
      \hline
      \hline
      1.0 &   &   &   & 0.539(4) & 1.0 \\
      \hline  
      ${1\over8}$ & ${1\over16}$ & ${1\over64}$ & ${1\over384}$ &  0.522(28) & 0.76(3)\\
      \hline
      0.25  &      &0.051&  & 0.564(34) & 0.79(2)  \\
      \hline
      0.4 & 0.12 &   &   &  0.563(31) & 0.88(2)   \\
      \hline
      0.8 & 0.06 &   &   &  0.563(26) & 0.83(3)   \\
      
      \hline
      \hline
    \end{tabular}
    \label{table:mrho}
  \end{table}
  \subsection{$L_s$ dependence of $m_{\rm res}$}
  \label{sec:decay}
\begin{figure}[t]
    \includegraphics[width=0.4\textwidth]{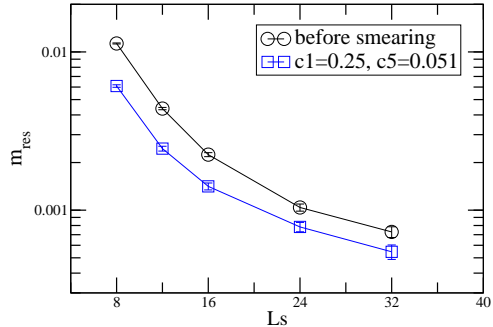}
    \vspace{-1pc}
    \caption{$L_s$ dependence of $m_{\rm res}$.}
    \label{fig:decay}
    \vspace{-1pc}
  \end{figure}

  Given that smearing has not removed the crossings at $-M_5 = -1.6$ (lower
  panel of Fig.~\ref{fig:sflow}), we have done further investigations on the
  decay of $m_{\rm res}$ with $L_s$.
  As shown in Figure~\ref{fig:decay}, $m_{\rm res}$ on the smeared lattices decays essentially the same as on the unsmeared lattices. 
  Contrary to the quenched DBW2 case \cite{ref:DBW2}, where the residual masses show
  simple exponential decay
  with $L_s$, here both smeared and unsmeared lattices show a more complicated
  behavior.
  The same phenomenon has also been seen in our quenched simulations with Wilson gauge action \cite{ref:quenchedDWF}, 
  where many topological dislocations are also observed. 
  
  \section{CONCLUSIONS}
  \label{sec:conclusion}
  We have found that for DWF, the smearings we have chosen do not change
  the lattice scale, but introduce additional mass renormalization factors.
  The smeared gauge fields are smoother, but we conclude that the smearings
  we have chosen do not change the topological dislocations present on
  the dynamical lattices, thus the large $L_s$ behavior of $m_{\rm res}$
  stays the same. Smearing does not appear to help DWF simulations at strong coupling. However, for $a^{-1} \sim 2$ GeV, where dislocations
  are suppressed due to weaker coupling, and a practical choice of $L_s$
  (say, 16), dynamical 3 flavor simulations with DWF are ready to do with
  current techniques.

\end{document}